\documentclass[a4paper,11pt]{article}
\usepackage{amsmath, amsfonts, amssymb}
\usepackage[cp1251]{inputenc}
\usepackage[dvips]{graphics}
\usepackage[russian]{babel}
\usepackage{epsfig}
\begin{document}

\title{\LARGE \bf
Multicolor electrophotometry of the peculiar object V1357 Cyg=Cyg
X-1 in the period 1986 -- 1992.}
\author{\bf A.N. Sazonov$^1$
\thanks{E-mail: sazonovaleksandr2010@yandex.ru}}

\date{\it  \small  $^1$ Sternberg State Astronomical Institute,
Lomonosov Moscow State University, Russia\\}

\date{\it  \small This paper is dedicated to Tatyana Musatova, who remain continuing
sources of inspiration\\}.

\maketitle
\renewcommand{\abstractname}{}
\begin {abstract}

\bf Abstract \rm Observations of close binary system (CBS) V1357
Cyg=Cyg X-1 in the $WBVR$ bands are reported. Photometry was
carried out at the telescopes located in Kazakhstan, Uzbekistan
and Crimea equipped with single-channel $WBVR$-photometer with
photon counting. 1358 individual observations in $WBVR$ bands on
202 nights were obtained.

The analysis of the photometric data allows to conclude that
different photometric effects are superimposed on the orbital
light curve of the system. Among them are brightness declines,
outbursts with different duration and amplitude, chaotic
variability, which sometimes exceeded the ellipsoidal variability.
Strong brightness decreases with magnitude $0^m.035 - 0^m.045$ are
observed, they are equal to the contribution of accretion disc to
the total luminosity of the system.

10 years of photometric monitoring allowed to detect continous
brightening of the system, followed by luminosity decline (slow
outburst). After the maximum in 1994-96 the decline was equal to
7\% in the $W$ band, 4\% in the $V$ band, and 2\% in the $V$ and
$R$ bands.

This paper continues the series of works on close systems with
X-ray sources.

\textbf{ Key words}: Close binary systems - luminosity -
ellipsoidal variability - accretion disc
\end{abstract}

\newpage
\section{INTRODUCTION}

Bright double X-ray systems have long series of electrophotometric
observations, which allow to determine many main physical
parameters of these CBS. But nevertheless many fundamental
properties of stellar matter flows in these systems remain
unclear.

The spectra of these objects contain nesessary additional
information. Taking into account the difficulties in studies of
such double systems, precision and detailed spectroscopic
observations should be carried out for the astrophysical objects
which have already been investigated by other methods.

The specialized new generation X-ray satellites EINSTEIN, ROSAT,
GINGA, ASCA, RXTE, CHANDRA, GRANAT, MIR-KVANT, GRO, XXM, and
others have systematically studied double systems with neutron
stars and black holes.

The close system V1357 Cyg=CygX-1 is the only one CBS which
satisfies all these selection criteria: it is one of the brightest
X-ray double systems in the optical wavelengths, it is studied in
detail by photometric methods, attracts attention of X-ray
observatories and is the first candidate for black holes (BH).

The relativistic object Cyg X-1 was discovered in X-rays and later
was identified with the optical companion of HDE 226868 with
$V$-band brightness of $\approx9^m$.

\subsection{GENERAL CHARACTERISTICS OF THE DOUBLE SYSTEM AND THE
TASKS OF INVESTIGATION}

Up to now this system remains No.1 black hole candidate, because
the mass of the relativistic component in this CBS exceeds
7M$_{\odot}$ ~\cite{Lyutyi1973}.

Recently investigations in different spectral bands resulted in
discovery of new fine photometric effects for Cyg X-1. For
example, besides orbital variability with period of $5^d.6$
different kinds of outbursts were revealed, and the so-called
precession period of $147^{d}/294^{d}$ was discovered.

Significant correlation between the long-period variations of
optical and X-ray (2 -- 10 keV) radiation is observed, typically
with time delay of one or two weeks.

According to generally accepted model the CBS consists of the
optical star -- supergiant of spectral class O9.7Ia,b ($V=8^m.8$)
and the relativistic object, which is surrounded by an accretion
disc (AD), radiating mainly in X-rays: $L_{x}= (3.3 -
5.5)\cdot10^{37}$ erg/s ~\cite{Dolan1979}. The distance to the
object is $\sim 2,5$ kpc. The temperature of the supergiant can be
estimated as 25000 K ~\cite{Bruevich1978}, 29500 K
\cite{Aab1984a}, \cite{Aab1984b} and up to 32000 K
~\cite{Canaliso1995}.

The contribution of AD to the total luminosity of the system is
only $4\%$  in the $V$ band
~\cite{Bruevich1978};~\cite{Voloshina1997}.

Cyg X-1 system has fast (about milliseconds) non-periodic
variability of X-ray radiation ~\cite{Oda1971}, which is one of
the important properties of matter accretion on the black hole
~\cite{Cherepashchuk1996};~\cite{Cherepashchuk2003}.

The classical effect of regular variability is observed for Cyg
X-1 system: the ellipsoidal effect, which results from tidal
deformation of the shape of the optical star, which has
deterministic character for such CBS.

The inclination, i, remains relatively uncertain. On the basis of
analysis of absorption lines,  ~\cite{Gies1986a} estimated
$i=33^\circ±5^\circ$.

On the other hand, the polarimetric measurements of
~\cite{Dolan1989} yield $25^\circ-67^\circ$. In our study, we
consider the inclination range of $30^\circ-60^\circ$.

We assume the black hole and the primary masses of 20 and
$40M_{\odot}$ ~\cite{Ziolkowski2005}, respectively, and a circular
orbit.

The radius of the primary is taken as
$r_{\star}=1.58\times10^{12}cm$ (\cite{Gies2003}).

The , and the orbital velocities are .

\bigskip

 \textbf{THE MAIN TASKS OF THIS WORK ARE}:

\begin{itemize}

\item to obtain homogeneous observational data in the $W$, $B$,
$V$, $R$ bands;

\item to obtain dense rows of photometric observations every
season;

\item to obtain observations with high temporal resolution to
reveal fast variability and fine photometric effects;

\item to obtain observations covering the moments of accretion
disc entering the eclipse and exit from eclipse.

\item to investigate the behaviour of the system at "off" and "on"
states on the base of photometric data.
\end{itemize}

\subsection{INTRODUCTORY NOTES FOR THE
OBSERVATIONS OF THE CLOSE SYSTEM}

It is nesessary to remind that at the beginning of 1980-ies
analysis of X-ray observations of the system at 30 keV band
~\cite{Manchanda1983} revealed the periodicity with characteristic
time of $\sim300^d$.

Sometimes later ~\cite{Predhorskiy1983} discovered long-time
variability of the system with period $294^d$ in the passband 3-6
keV using uniform observations from X-ray satellite Vela. Later
~\cite{Kemp1983} has shown the existence of this periodicity in
optical wavelengths and concluded that the shape of the optical
light curve is different at various phases of the $294^d$-long
precession period.

The orbital light curve during outburst differs from the average
annual light curve by extra extinction of radiation at orbital
phases $\varphi =0.00 - 0.50$, this effect can last about 40 days
after the end of outburst, as it happened, for example, in 1980
~\cite{Brocksopp1999a}; ~\cite{Voloshina1997}.

\section{OBSERVATIONS OF THE CLOSE SYSTEM}

Photoelectric observations of V1357 Cyg=Cyg X-1 were carried out
in the $WBVR$ system. Total number of observations is 1358
individual estimates on 202 nights in the period 1986 -- 1992.

The observations were carried out at the reflectors Zeiss-600,
Zeiss-1000 and AZT-14 with the same equipment: single-channell
photometer with $WBVR$ filters and FEU-79 photomultiplier
(multialkali photocathode S-20).

For each season the coefficients of transformation of instrumental
photometric system were determined.

The photometric observations are presented in Table 1. Reduction
was carried out with differential method. Sometimes the extinction
coefficients were measured by Nikonov's method, usually during the
best photometric nights. Then the mean coefficients $\xi$ were
calculated for each season of observations, and for each night the
zero-points $\eta$ were computed using the mean values of $\xi$.
Mean values of $\xi$ and their errors are presented in Table 2,
where n is the number of nights used for calculation.

Table 1 with photometric data looks this way:

\begin{table}[htbp]
\caption{Results of observations}
\begin{tabular} {|p{22pt}|p{80pt}|p{40pt}|p{35pt}|p{35pt}|p{35pt}|p{35pt}|p{20pt}|}
\hline \No \par No/No &JD2400000+.& $\varphi$&
$W$& $B$ & $V$ & $R$ & n \\
\hline 1& 46615.3830 & 0.873 &  15.075 &  15.125
& 14.667 & 14.328 & 06 \\
\hline 2& 46616.4574 & 0.983 &  15.144 &  15.314
& 14.946 & 14.660 & 08 \\
\hline 3& 46618.4099 & 0.181 &  14.855 &  15.306
& 14.779 & 14.585 & 06 \\
\hline 4& ....... & ........ &  ...... &  ......
& ...... & ...... & .. \\
\hline
\end{tabular}
\label{tab1}
\end{table}

where n is the number of individual observations in a night.

Table 2 with extincion and transformation coefficients is reported
in the Database.

\begin{table}[htbp]
\caption{Transformation coefficients}
\begin{tabular}{|p{4cm}|p{4cm}|p{4cm}|}
\hline 1. Season of observations: July - September 1986, 1987,
1988. Reflector: AZT-14 (400 mm) Tjan-Shan observatory & 2. Season
of observations: August - October 1986, 1987, 1988, 1989, 1990,
1992. Reflector: Zeiss-600 Crimean Observatory SAI; & 3. Season of
observations: July - September 1986, 1987, 1988, 1990, 1992,
1994. Reflector: Zeiss-600 Maidanak Observatory SAI; \\
\hline \vspace{3pt}
\parbox{4cm}{
$\xi_{V} =0.054 \pm 0.002 $ \par $\xi_{W-B}= 0.997\pm 0.009$ \par
$\xi_{B-V} =0.929\pm 0.005$ \par $\xi_{V-R}=1.068\pm 0.008$  \par
$n=27$} & \vspace{3pt}
\parbox{4cm}{
$\xi_{V} =0.013\pm 0.003;$  \par $\xi_{W-B}= 0.962\pm 0.005;$\par
$\xi_{B-V} =1.102\pm0.003;$ \par $\xi_{V-R}=1.088\pm 0.004;$ \par
$n=38$} & \vspace{3pt}
\parbox{4cm}{
$\xi_{V} =0.012\pm0.003;$   \par $\xi_{W-B}=0.958\pm0.004;$  \par
$\xi_{B-V}=0.937\pm0.007;$  \par $\xi_{V-R}=1.065\pm0.007;$  \par
$n=41$ \strut}\\
\hline
\end{tabular}
\end{table}

\subsection*{ELECTRONIC ADDRESS OF THE DATABASE
OF PHOTOMETRIC OBSERVATIONS}

The results of observations of CBS V1357 Cyg=Cyg X-1 in the
optical wavelengths are presented in Table 1 at the address:
{http://lnfm1.sai.msu.ru/$\sim$ sazonov/~Cyg~X-1}

\subsection*{COMPARISON STARS AND THE CONTROL STAR}

The observations of the program object were made relative to the
comparison stars and the control star from the list presented in
Table 3:

\begin{table}[htbp]
\caption{Comparison and control stars}
\begin{tabular}{|p{17pt}|p{25pt}|p{55pt}|p{25pt}|p{35pt}|p{35pt}|p{35pt}|}
\hline \No \par no/no &Star& $BD/GCVS$&$V$& $U-B$ &$B-V$ & $V-R$ \\
\hline 1& c & +34:3812 & 9.082 & -0.013 & -0.029
& 0.024  \\
\hline 2& a & +34:3816 & - & - & -
& -      \\
\hline 3& 1 & +35:3895 & 6.998& +0.051*& 0.040
& 0.031  \\
\hline
\end{tabular}
\end{table}

Note:*- color $(W - B)$ is reported.
\bigskip

Orbital phases of CBS V1357 Cyg=Cyg X-1 are presented according to
the ephemerides by ~\cite{Brocksopp1999a},~\cite{Brocksopp1999b}:

\bigskip
Min I hel = JD 2441163.529$(\pm 0.009)$ + $5^d.59985(\pm 0.00012)$

\bigskip
The phases of precession periods were computed according to the
ephemerides by ~\cite{Karitskaya2001}, ~\cite{Kemp1987}:

\bigskip
Min I hel = JD 2449953 + $147^d.0\cdot$E

\bigskip
Min I hel = JD 2449953 + $294^d.0\cdot$E
\bigskip

According to these elements the minimum at the orbital phase
$\varphi$ =0.0 corresponds to location of X-ray source of CBS in
front of the optical component (the moment of upper conjunction of
the X-ray source), and the mimimum at the orbital phase $\varphi$
=0.5 occurs when the X-ray source is behind the normal star.

The observations in the period 1986--1987 were carried out
relative to the standard star BD+34:3816, which was later replaced
by BD+34:3812, because BD+34:3816 was suspected to be variable
with small amplitude ~\cite{Walker1978a}, ~\cite{Walker1978b}.
Later ~\cite{Goranskij1999} confirmed that star BD+34:3816 is a
variable with small amplitude ($\sim 0^m.01$) and periodical
components in the light curve. The result was based on special
photometric investigation.

The star BD+35:3895 was used as control star during these
observations.

It is nesessary to stress that in 1988--1992 author used the star
BD +34:3812 as local standard because variable V1357 Cyg and BD
+34:3812 have comparable brightness. Besides, BD +34:3812 is the
star of early spectral class, but it's color is different from
V1357 because of strong interstellar extinction. The difference in
color results in significant corrections which were applied during
reduction.

\section{REGULAR VARIABILITY OF BRIGHTNESS OF CLOSE SYSTEM}

While analysing the light curves in the period 1986--1999 the
strong physical irregular variability of V1357 Cyg attracts
special attention. It is superposed on the periodical brightness
changes, resulting from ellipsoidality, and depending also on the
precession period $147^d/294^d$ (Fig.1-4).

\bigskip
See (Fig.5) $B$ from $V$, vs. orbital phase $\varphi$ for the
1989-1999 season.
\bigskip

It is nesessary to note that during the constant increase of
amplitude of variability in all four spectral bands every year of
observations the irregular brightness fluctuations take place in
phase with regular changes, and while decrease of amplitude in
antiphase ~\cite{Lyutyi1974}.

The author used his 7-year long multicolor $WBVR$ observations to
investigate the photometric variability of optical component of
CBS. For qualitive analysis of these data the author used also the
published optical ~\cite{Kemp1978}; ~\cite{Kemp1987};
~\cite{Lyutyi1985}; ~\cite{Voloshina1997} and X-ray
~\cite{Zhang1996a}; ~\cite{Zhang1996b}; ~\cite{Cui1996} data.

The errors of these observations are typically $0^m.003 - 0^m.007$
on different nights.

The figures show that the mean light curve of the system V1357
Cyg=Cyg X-1 is close to sinusoidal, although it is evident that
the minimum at phase $\varphi=0.50$ is slightly wider and
shallower than at orbital phase $\varphi=0.00$.

The orbital phase $\varphi=0.00$ corresponds to the moment of
upper conjunction of X-ray source and the optical component of the
system.

This observable difference at the light curves can be interpreted
as the evidence that the optical component fills it's Roche lobe,
and the star is not ellipsoidal, but pear-shaped. In this case the
role of gravitational darkening is increasing.

The observations revealed flares, which result from appearance of
temporarily hot regions, so-called "hot spots"{}. The analysis of
X-ray emission reveals strong correlation between appearance of
"hot spots"{} and decrease of X-rays ("dips"{}), the reason is
perhaps the emergence of absorbing gas at the light beam of the
observer.

The non-stationarity of gas flow through the inner Lagrange point
$L_1$ from the optical component of the system to the accretion
disc (AD) of the relativistic object can generate shock waves in
the gas which surrounds the system. The shock waves are formed
also when gas enters the AD, causing redistribution of scattering
and absorbing matter in the vicinity of the relativistic object,
and this fact is revealed in the optical radiation.

Different short-term variability is occasionally superposed on the
mean orbital light curve of the system, it is not strictly
dependant on the orbital and precession phases. The flare
variability seems to be integrated to this occasional irregular
variability: strong and long outbursts with amplitude up to
$\Delta=0^m.03 - 0^m.04$, and also short-period flares with
amplitude of $\Delta=0^m.01$; long (up to 7--10 days) brightness
decreases with amplitude up to $\Delta=0^m.040 - 0^m.045$; and
also chaotic irregular variability.

The mentioned above types of variability, as a rule, coincide with
local maxima and minima of X-ray radiation.


This type of variability was detected during studies of
photometric changes on the light curve with characterisic times
from 2-3 days to 10-12 days (as in the seasons 1986 and 1988; see
Fig. 1,3).


All these photometric phenomena are probably due to the activity
of X-ray component during Cyg X-1 transition to the "soft" state
~\cite{Zhang1996a}; ~\cite{Zhang1996b}; ~\cite{Cui1996};
~\cite{Brocksopp1999a}.

Ellipsoidal light curve shows relatively short outbursts with
amplitude up to $\Delta=0^m.020\pm0^m.030$ and length of about 3-4
hours during the night of observations. At these observing seasons
the symmetry of double wave is absent at orbital phases
$\varphi=0.00$ and $\varphi=0.50$. This feature of the light
curves of the optical component of V1357 Cyg is probably connected
to some asymmetry of strong gas flows in the system.

\section{FAST VARIABILITY OF THE SYSTEM}

Fast variability of the star in $WBVR$ bands was investigated at
minimum and maximum brightness on time spans of $\sim $ 60-90 s.
The deviations of brightness from the mean value within $3\sigma$
on time intervals of 60-90 s were detected. The amplitude of fast
variability is up to 0$^{m}$.005 $\div$ 0$^{m}$.007. Such fast
variability is probably characteristic of short-time outbursts in
the system.

Analysis of fast variability and functional dependence between
optical and X-ray bands revealed very low dependence at the level
of a few percent.

\section{ANALYSIS OF THE CHANGE OF THE MEAN LEVEL OF BRIGHTNESS
OF THE SYSTEM}

Qualituve comparison with the observations of 1986 -- 1998 allows
to conclude that long-term small-amplitude "outburst" took place
at the system. This long-term variability is probably connected to
the evolutionary changes of supergiant in the close system. This
is confirmed by the change of color indices at there epochs.
Consecutive spectral observations revealed the cooling of
supergiant by about 2000 K by comparison to the "peak temperature"
of 1995 -- 1996. The spectral type of the supergiant also changed.
The color excess also slightly decreased compared to the previous
value of $E(B-V)=1.12$ ~\cite{Walborn1973}.

The annual changes of the mean brightness of close system V1357
Cyg=HDE 226868=Cyg X-1 for the period 1986 -- 1998 are presented
below.

\begin{table}[htbp]
\caption{The data on annual changes of the mean brightness of
V1357 Cyg=Cyg X - 1}
\begin{center}
\begin{tabular}{|p{19pt}|p{25pt}|p{25pt}|p{25pt}|p{25pt}|p{25pt}|p{25pt}|p{25pt}|p{25pt}|p{25pt}|p{15pt}|}
\hline \No \par no/no &Year& $W$&$\sigma$& $B$ &$\sigma$ & $V$& $\sigma$ & $R$& $\sigma$& n \\
\hline 1& 1986 & 9.384 & 0.009& 9.678& 0.006
& 8.854 & 0.004& 7.956 & 0.002& 484 \\
\hline 2& 1987 & 9.373 & 0.008& 9.663& 0.007
& 8.849 & 0.005& 7.941 & 0.007& 270 \\
\hline 3& 1988 & 9.362 & 0.003& 9.659& 0.002
& 8.844 & 0.008& 7.938 & 0.006& 604 \\
\hline 4& 1989 & 9.354 & 0.006& 9.651& 0.003
& 8.840 & 0.008& 7.935 & 0.002& 358 \\
\hline 5& 1990 & 9.341 & 0.004& 9.642& 0.003
& 8.837 & 0.007& 7.930 & 0.002& 143 \\
\hline 6& 1991 & 9.330 & 0.006& 9.635& 0.004
& 8.833 & 0.002& 7.928 & 0.008& 158 \\
\hline 7& 1992 & 9.321 & 0.006& 9.628& 0.003
& 8.831 & 0.003& 7.926 & 0.006& 168 \\
\hline 8& 1993 & 9.308 & 0.009& 9.623& 0.008
& 8.830 & 0.006& 7.924 & 0.003& 186 \\
\hline 9& 1994 & 9.298 & 0.008& 9.621& 0.006
& 8.829 & 0.003& 7.922 & 0.002& 195 \\
\hline 10&1995 & 9.290 & 0.012& 9.619& 0.008
& 8.827 & 0.006& 7.920 & 0.004& 208 \\
\hline 11&1996 & 9.292 & 0.004& 9.623& 0.002
& 8.832 & 0.007& 7.921 & 0.006& 216 \\
\hline 12&1997 & 9.296 & 0.008& 9.627& 0.006
& 8.836 & 0.005& 7.923 & 0.004& 218 \\
\hline 13&1998 & 9.301 & 0.007& 9.635& 0.004
& 8.840 & 0.005& 7.925 & 0.004& 228 \\
\hline 14&1999 & 9.317 & 0.010& 9.641& 0.006
& 8.843 & 0.005& 7.929 & 0.004& 241 \\
\hline
\end{tabular}
\end{center}
\label{tab1}
\end{table}

It is nesessary to note that the mean points for separate nights
of observations near orbital phases $\varphi = 0.0$ and $\varphi
=0.5$ sometimes deviate from the light curve of the system. This
fact can probably be explained by small mutual partial eclipses of
the optical component of CBS and non-stationary accretion
structure (AD and, probably, the gas crown of the system),
localized in the close vicinity of the invisible relativistic
companion.

\section{QUALITATIVE ANALYSIS OF THE OPTICAL AND X-RAY
DATA FOR THE CLOSE SYSTEM}

Qualitative analysis of the optical and X-ray data for the system
for the period from 1986 until 1993 (author's data) and from 1994
until 2005 (published data from other authors) was carried out
(See Fig.6).

Specialized X-ray observatories EXOSAT, RXTE/ASM, RXTE/PCA, BATSE
and others gave extensive data on X-ray emission of Cyg X-1 for
analysis and comparison with the optical data.

While averaging the points of optical observations for the period
1995 -- 1996 (peak of active state of the system) and up to 2005
and comparing them to the X-ray data obtained by RXTE/ASM at the
same period, we can note the increase of X-ray activity of the
system with simultaneous decrease of the mean optical brightness.

Such behaviour of the system can be explained in the frameworks of
the selected gas-dynamical model: some change (in this case
increase) of accretion rate with simultaneous increase of the
radius of the optical component of the system (in the frameworks
of Roche model -- the change of degree of filling by optical star
of its Roche lobe). The temperature of supergiant decteases by
some value.

To reveal the degree of correlation between optical and X-ray
radiation it is nesessary to take into consideration the existing
connection from the frequency analysis: in the optical and X-ray
light curves the orbital components with significant amplitudes
are present. The amplitude in the X-ray component of the spectra
is about 0$^{m}$.16, and in the optical band -- 0$^{m}$.045
(Belloni, Mendez et al., 1996). Analysis of fast variability and
functional dependence between optical and X-ray bands revealed
very low dependence at the level of a few percent, while
comparison of long-term changes in these bands gives significant
correlation of $42-45 \%$ ~\cite{Brocksopp1999a};
~\cite{Brocksopp1999b}.

The correlation coefficients should be small because of the
changes of mean phase light curves of the system: double wave with
equal minima in the optics and single wave with narrow secondary
minimum in X-rays ~\cite{Brocksopp1999a}.

\section{REALISTIC MODEL OF SMALL AMPLITUDE LONG TERM OUTBURST}

{\bf Dynamo model to account for the behaviour of the peculair
object V1357 Cyg in the close binary system V1357=Cyg X-1.}

It is generally assumed that the physical nature of the solar
cycle is related to the action of a dynamo mechanism in the depths
of the solar convection zone.

\bigskip
По видимому, такой же механизм динамо действует и в глубокой, но
тонкой конвективной зоне сверхгиганта, каковым является оптический
компонент в ТДС V1357 Cyg= Cyg X-1.
\bigskip

This mechanism operates due to the combined effect of differential
rotation and the so-called $\alpha-effect$, which is related to
convection breaking the mirror symmetry in the rotating body.

This mechanism is probably also responsible for the evolution of
the magnetic field in other celestial bodies, including the Earth,
stars and galaxies ~\cite{Nefedov2010}.

The dynamo effect is a process by which a magnetic field is
generated by the flow of an electrically conducting fluid. It is
believed to be responsible for magnetic fields of planets, stars
and galaxies ~\cite{Moffatt1978}. Fluid dynamos have been observed
in laboratory experiments in Karlsruhe ~\cite{Stieglitz2001} and
Riga ~\cite{Gailitis2001}. More recently, the VKS experiment
displayed self-generation in a less constrained geometry, i.e., a
von Karman swirling flow generated between two counter-rotating
disks in a cylinder~\cite{Monchaux2007}, ~\cite{Christophe2009}.

We suppose that the described above event of slow outburst: the
increase of brightness starting at about 1986 with probable
maximum in the period 1994-1995 and further decline to the usual
level at 2000, should be considered as the process of stellar
activity of the normal star (supergiant) in the double system.

We suggest that the physical nature of this event is to some
extent similar to the solar cycle, that is somehow connected to
changes of the large-scale magnetic field of the supergiant. This
proposal is based on the fact that the temporal and energetical
characteristics of the slow outburst differ significantly from
such parameters as the time of X-ray outbursts (milliseconds),
connected with the accretion of matter on the black hole, or from
the period of the orbital variability (5$^d$.6).

Let us consider the known processes leading to formation and
evolution of magnetic field on normal stars. There might be some
variant of stellar dynamo acting in the convective envelope of the
supergiant.

According to modern ideas the supergiants really possess such
convective envelopes, with size about of the radius of the star.

The radius of the primary is taken as
$r_{\star}=1.58\times10^{12}cm$ (~\cite{Gies2003}).


What is known about convection, are there some ideas about the
degree of differentiality of the rotation?

Let us consider the action of stellar dynamo in the spherical
envelope in the framework of a simpified model. The differential
rotation acts on the poloidal magnetic field and creates toroidal
magnetic field. From other side, the global rotation of the star
leads to mirror asymmetry of convection, which allows to create
poloidal magnetic field from toroidal field (so called
alpha-effect) and to close the chain of self-excitation of
magnetic field. This scheme was proposed by Parker in 1955 and is
called Parker dynamo.

We will try to estimate the effectiveness of Parker's dynamo in
supergiants on the basis of known facts about convection in
supergiants.

It is well known that Parker's dynamo may excite the waves of
large-scale quasistationary magnetic field in the spherical
envelope (so called dynamo waves) running from medium latitudes to
the equator, or, if the sign of alpha-effect is opposite, from
medium latitudes to the pole. These waves are usually close to
harmonical, because toroidal magnetic field changes proportionally
to $\sin(t/t_c)$, where $t_c$ is the cycle period.

At present we do not have direct evidence that the observed slow
outburst is a periodical (or nearly periodical) event, but this
proposal seems plausible. In this case it turns out that the
photoelectic sygnal, showing the slow outburst of the supergiant
in the system Cyg X-1, is different from harmonical.

For quantitative characterisation of this difference we introduce
the duty cycle $Q=T_{-}/T_{+}$, where $T_{-}$ is the time in low
state, and $T_{+}$ is the time in high state. Taking $T_{+}=14$
years (from 1986 until 2000) and $T_{-}=27$ years (from 1970 until
1986 and from 2000 until 2011) we obtain $Q=1.9$. Assuming that
for sinusoidal sygnal the moment of transition from low state to
high state is at $\sin(t/t_c)=0.5$ we get $Q=0.5$.

So, the possibility of interpretation of the observations
discussed above depends on the possibility for Parker's dynamo to
produce periodically changing magnetic fields with high duty
cycle. Moss et al. ~\cite{Moss2008} noted that with definite set
of directing parameters Parker's dynamo really produces
periodically changing magnetic field differing from sinusoidal.
They proposed to search the stars with such behaviour of magnetic
field among magnetically active stars, but they did not
characterise its behaviour by duty cycle.

Periodically changing magnetic fields with high duty cycle, so
called dynamo-splashes, were discovered experimentally while
studying dynamo in the laboratory experiment VKS in Lyon, France.

\bigskip

Previous simulations, using the mean flow (time averaged) of the
VKS experiment or an analytical velocity field with the same
geometry, predicted an equatorial dipole: ~\cite{Marie2003},
~\cite{Bourgoin2004}, ~\cite{Gissinger2008}, ~\cite{Stefani2006}
in contradiction with the axial dipole observed in the experiment
~\cite{Christophe2009}.

\bigskip

As at present it is technically impossible to reproduce Parker's
dynamo in the laboratory and the self-excitation of magnetic field
in VKS experiment differs significantly from the stellar dynamo,
we give below a model for dynamo-splashes. We propose it as a
model for the occurrence of a slow outburst on the supergiant in
the system Cyg X-1 in the form, more adequate to Parker's dynamo.
At the same time we note that this model is close to the one
suggested by Fove:

\bigskip
Reference for VKS  ~\cite{Berhanu2007}, ~\cite{Petrelis2009},
~\cite{Moss2008}
\bigskip

  It seems likely that the excited magnetic field is, in some sense,
simply organized and can be described using relatively few
parameters. Therefore, a qualitative description can be achieved
by replacing the dynamo equations with an appropriately chosen
dynamical system of a moderately high order.

Our knowledge on the distribution of sources of magnetic field
generation (differential rotation and spirality) is limited, that
is why we simpify stellar dynamo equations to the corresponding
level. Following Parker (1955) ~\cite{Parker1955} we consider this
field to be axis-symmetrical and will average it over the radial
size of the convective zone.

This problem was first formulated in ~\cite{Ruzmaikin1981} (see
also ~\cite{Zeldovich2006}), where the following dynamical system
was suggested:

Then we will decompose the magnetic field to Fourier series on the
remaining variable (latitude) and will conserve minimal number of
Fourier modes, sufficient for the model to reproduce growing
magnetic field with non-vanishing magnetic moment. We also suppose
that growing and oscillating magnetic field is stabilized by the
simple mechanism for suppression of alpha-effect.

Then we obtain the following dynamic system (Nefedov and Sokolov,
2010 ~\cite{Nefedov2010}):

Basic MHD equations and the description of variables:

\begin{equation}
\frac{dA}{dt}=-A+\sigma^{.}DB-CB,
\end{equation}

\begin{equation}
\frac{dB}{dt}=-\sigma^{.}B+\sigma^{.}A,
\end{equation}

\begin{equation}
\frac{dC}{dt}=-\upsilon^{.}C+AB,
\end{equation}

\bigskip
\bigskip

Here, A corresponds to the toroidal component of the magnetic
potential and, to the poloidal magnetic field, B to the toroidal
component of the magnetic field, D to the dynamo number, and C to
the $\alpha-effect$. The physical meaning of the coefficient
$\sigma$ required addition clarification.

This system can reproduce dynamo-splashes within definite interval
of the parameters.

\bigskip

\section{Suggested model for the explanation of the slow and
small-amplitude outburst in the photospere of supergiant. Proposed
by A.N.Sazonov}

For the interpretation of the slow and small-amplitude outburst in
the photosphere of supergiant V1357 Cyg it is necessary to put
forward the following realistic physical suggestions:

1. Priming magnetic field is necessary for the action of dynamo
mechanism, this field should be further intensified and rebuild.

2. Close binary systems are fast-rotating stars, which should
generate strong magnetic dynamo. We should expect that processes
of reconnection are taking place, which are initiated by rotation
stress ~\cite{Jeffries1993}.

3. If an accretion disk is present is the system, it possesses its
own gas corona, which may be the source of disk (impulse)
outbursts.

4. The outbursts of different duration and amplitude in double
stellar systems are not analogous to the solar flares; it is
necessary to take into account the presence of strong stellar
coronas, which can serve as dominating mechanism for energy
storing ~\cite{Levy1989}.

5. Stellar outbursts are much more powerful than the solar flares,
perhaps because they take place on stars with stronger magnetic
dynamo.

6. Sufficiently fast rotation, combied with deep convective zone
may be present at supergiants similar to V1357 Cyg and increases
the dynamo activity on these stars.

7. Another remarkable feature of these close binary systems,
stellar microquasars, is the presence of jets and accretion disks
around relativistic objects.

8. The separation then corresponds to
$a=2.28r_{*}=3.60\times10^{12}$ cm, and the orbital velocities are
$3.13\times10^{7}$ and $1.56\times10^{7} cm s^{-1}$ for the black
hole and the primary, respectively. We hereafter use a  as the
unit, as it gives a measure of the asymmetry in the photon paths
during the orbital motion. The distance is assumed to be $D=2$ kpc
~\cite{Ziolkowski2005}.

\bigskip
\subsection{ Interpretation of the outburst}

All the assumptions we made give realistic basis for the
interpretation of long-term small-amplitude outburst, discovered
from photometric data by A.N.Sazonov

Thus, we conclude that our dynamical system can reproduce the
phenomenon of cyclic activity, including at least some chaotic
elements of this regime.

\section{CONCLUSIONS}

1. The light curve of the system for the period 1986 -- 1992 is in
good accordance within the errors to the mean light curves of the
other authors with their sets of observations. In this work
homogeneous and high-precision set of observations in the $W$,
$B$, $V$, $R$ filters is presented, which makes them unique and
valuable for obtaining additional information of the physical
proprties of the system. They allow to carry out analysis of fine
effects and derive information on other physical parameters of the
unique system.

2. Brighness decreases of $0^m.035 - 0^m.045$ are observed, which
is equal to the contribution of accretion disc to the total
luminosity of the system.

3. 10 years of photoelectical observations of the object reveal
annual increase, followed by the decrease of the mean brightness
of the system ("slow outburst"). The decline of brightness after
the maximum in 1994 -- 1996 was equal to $7\%$ in $W$ band, $4\%$
in $B$, in the bands $V$ and $R$ it was nearly $2\%$.

4. In general, we conclude that the constructed dynamical system
can qualitatively reproduce the temporal behavior of real dynamo
systems of various celestial bodies.

\section{ACKNOWLEDGEMENTS}

The author is thankful to D.Yu.Tsvetkov for the translation of the
paper and critical comments.

This paper is dedicated to Tatyana Musatofa, who remain continuing
sources of inspiration.

\begin{figure*}[!b]
\includegraphics[scale=0.9]{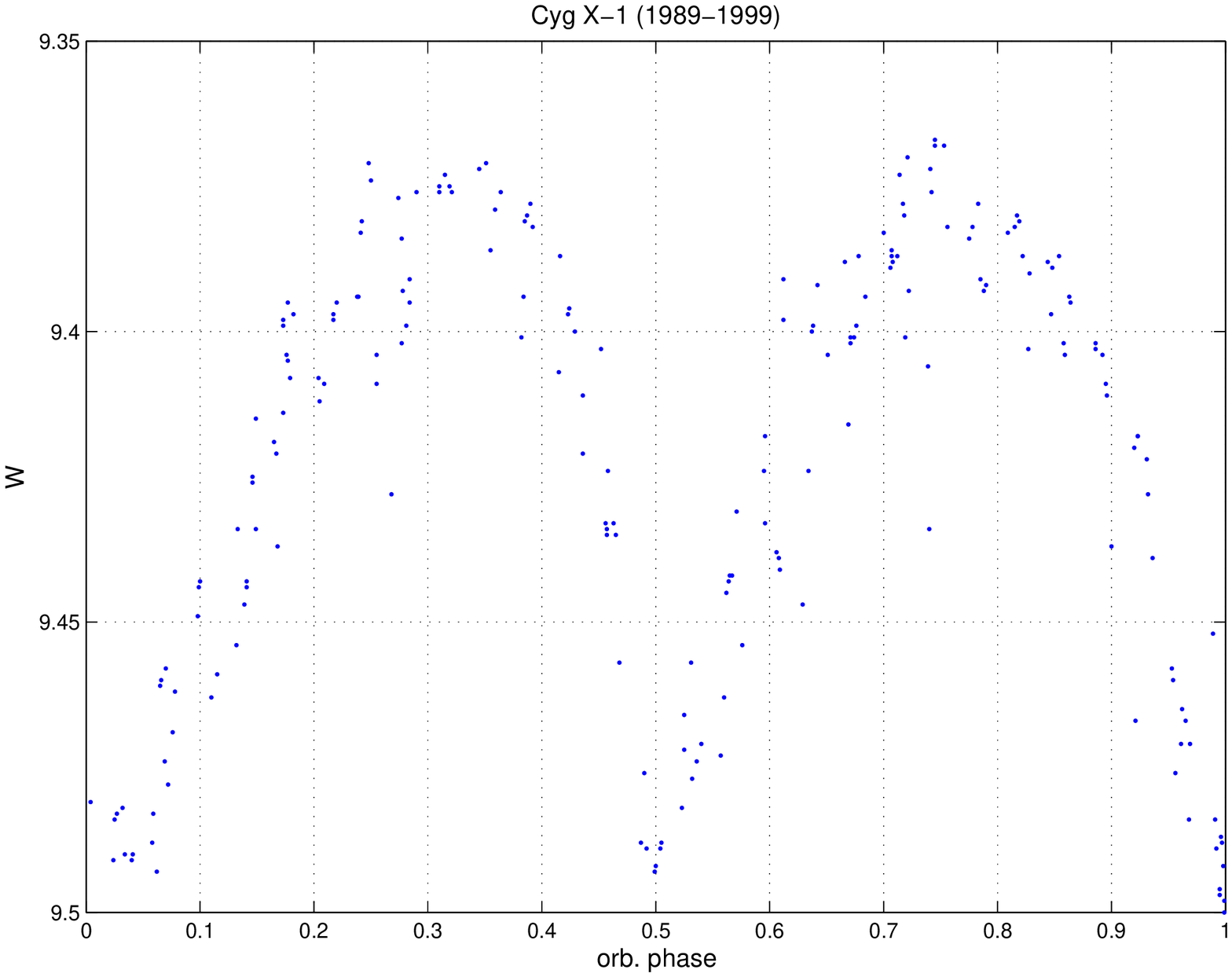}
\caption{$W$,vs. orbital phase $\varphi$ for the 1989-1999 season.
\hfill}
\end{figure*}

\begin{figure*}[p!]
\includegraphics[scale=0.9]{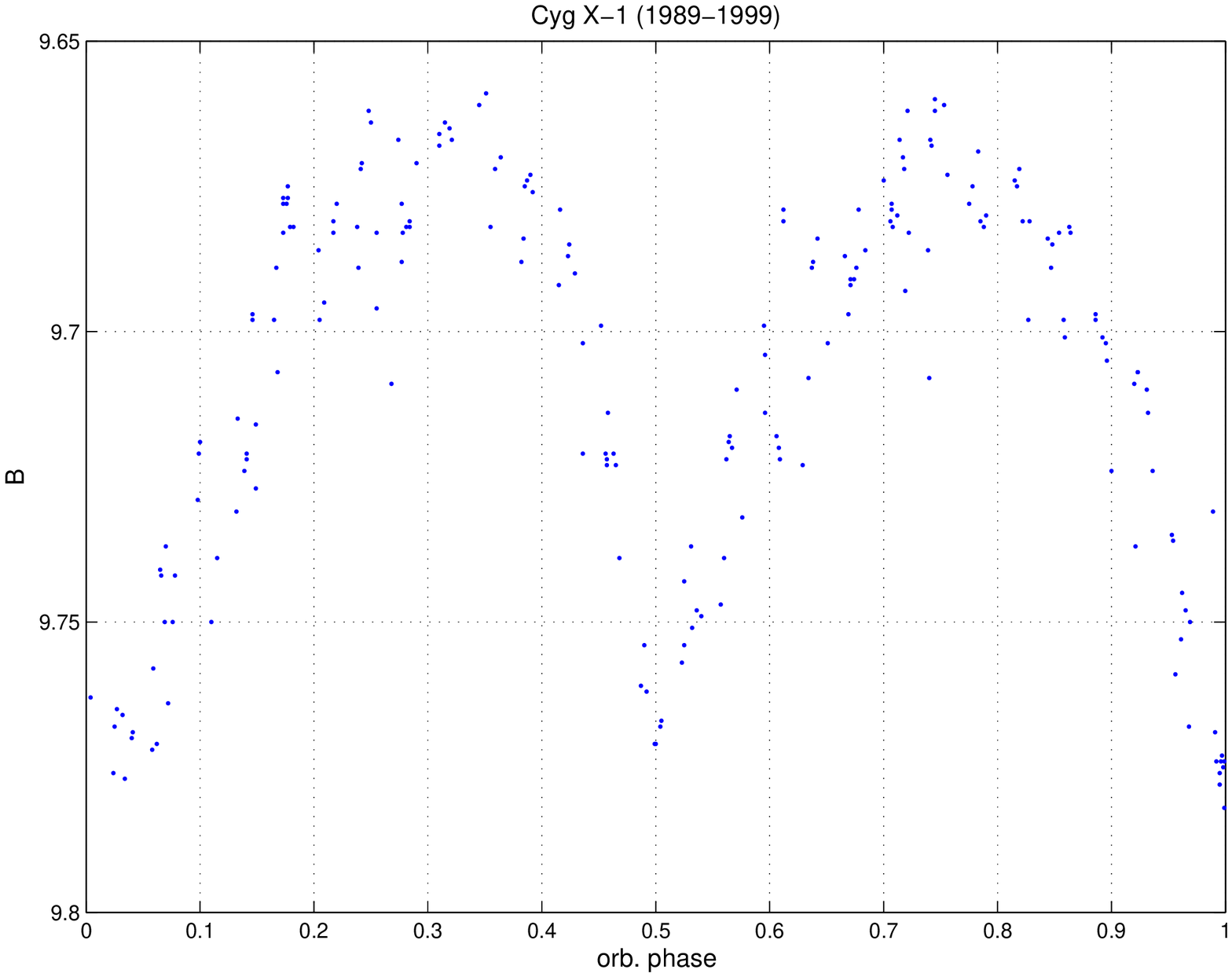}
\caption{$B$,vs. orbital phase $\varphi$ for the 1989-1999 season.
\hfill}
\end{figure*}

\begin{figure*}[p!]
\includegraphics[scale=0.9]{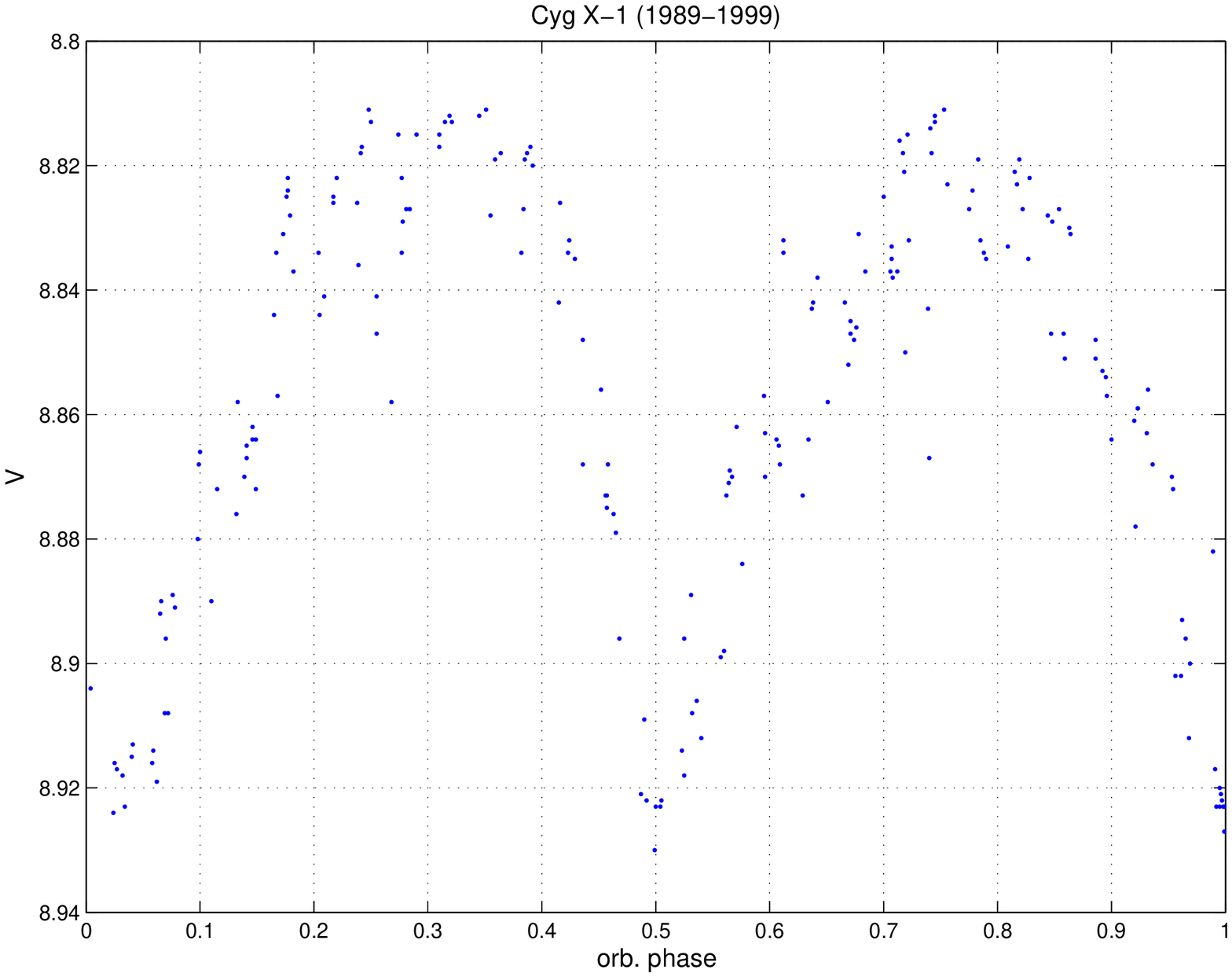}
\caption{$V$,vs. orbital phase $\varphi$ for the 1989-1999 season.
\hfill}
\end{figure*}

\begin{figure*}[p!]
\includegraphics[scale=0.9]{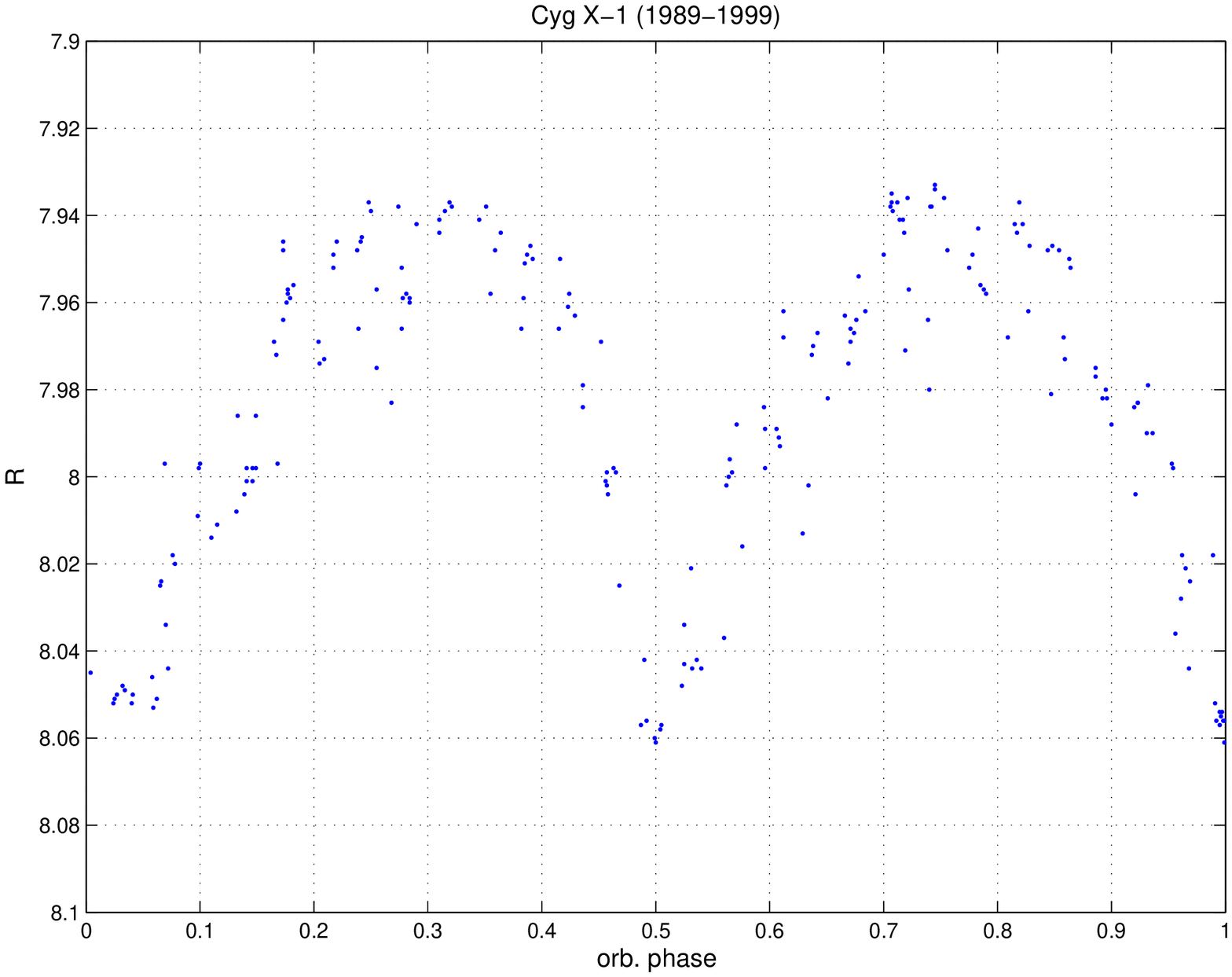}
\caption{$R$,vs. orbital phase $\varphi$ for the 1989-1999 season.
\hfill}
\end{figure*}

\begin{figure*}[p!]
\includegraphics[scale=0.9]{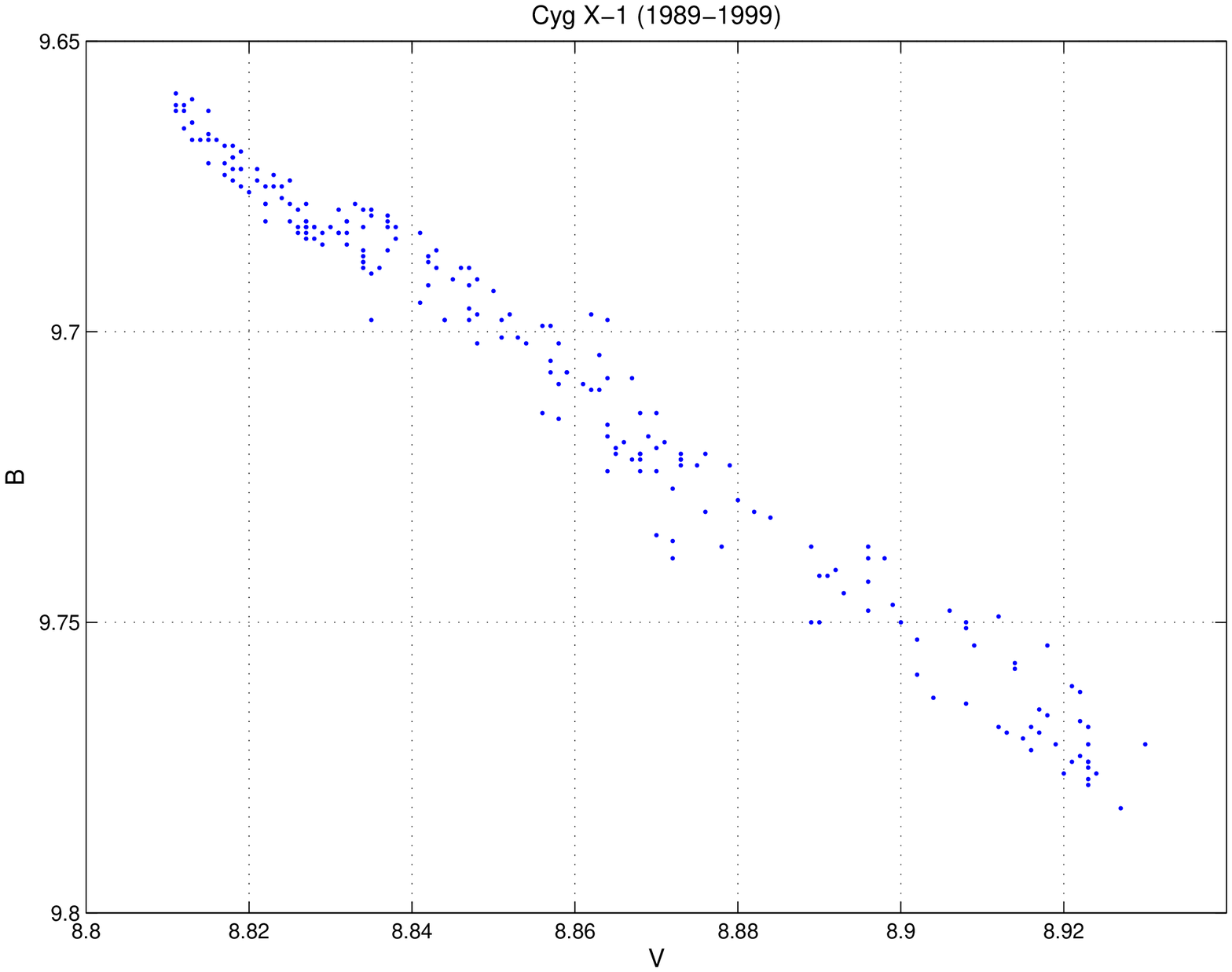}
\caption{$B$ from $V$, vs. orbital phase $\varphi$ for the
1989-1999 season. \hfill}
\end{figure*}

\begin{figure*}[p!]
\includegraphics[scale=0.9]{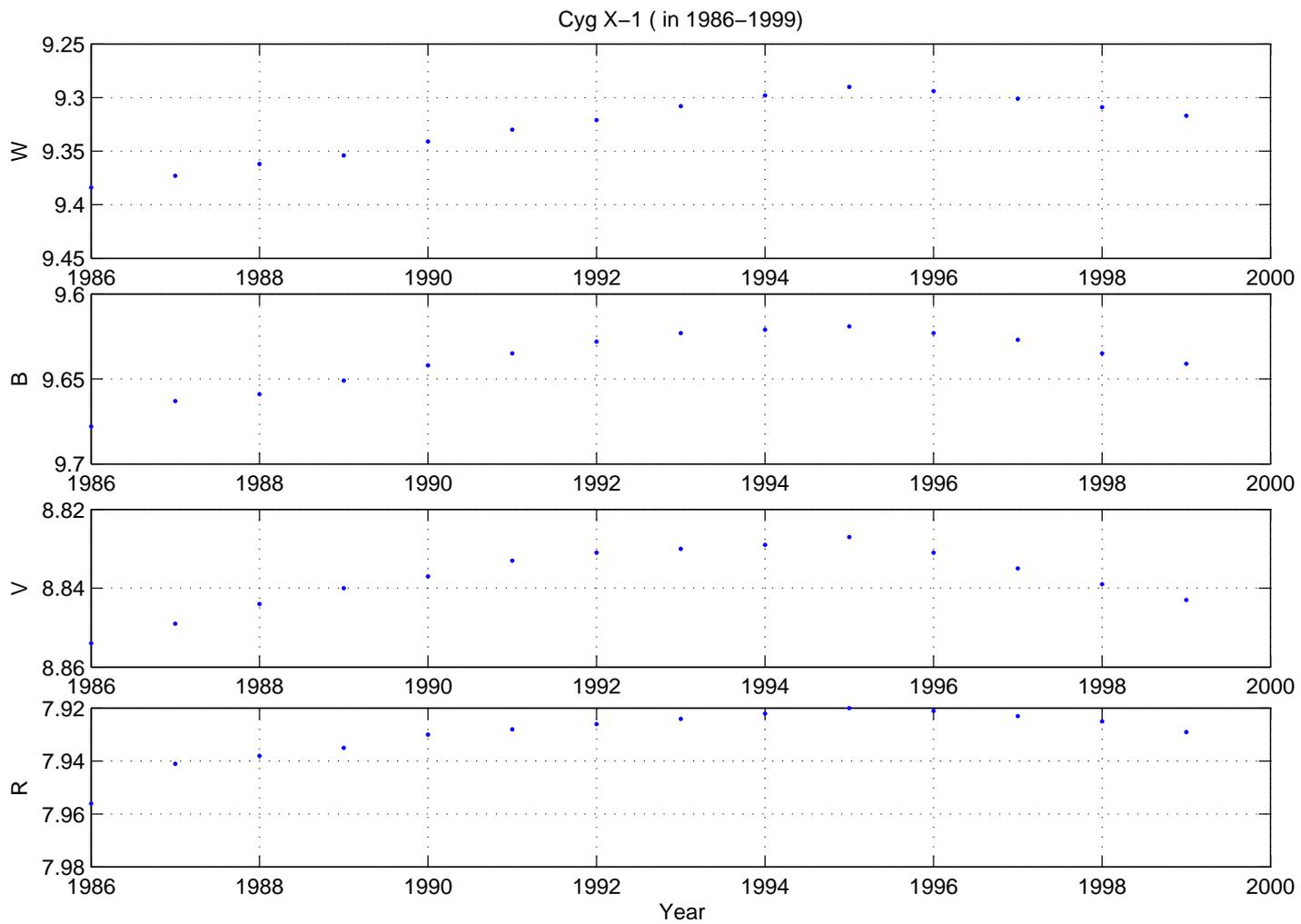}
\caption{$WBVR$,vs. orbital phase $\varphi$ for the 1986-2000
season. \hfill}
\end{figure*}

\end{document}